# 3D DNA origami crystals

*Tao Zhang, Caroline Hartl, Stefan Fischer, Kilian Frank, Philipp Nickels, Amelie Heuer-Jungemann, Bert Nickel, Tim Liedl**

*Faculty of Physics and Center for Nanoscience (CeNS), Ludwig-Maximilians-Universität, Geschwister-Scholl-Platz 1, 80539 München, Germany*

**Email: tim.liedl@physik.lmu.de*

**Engineering shape and interactions of nanoscopic building blocks allows for the assembly of rationally designed macroscopic three-dimensional (3D) materials with spatial accuracy inaccessible to top-down fabrication methods[1–3]. Owing to its sequence-specific interaction, DNA is often used as selective binder to connect metallic nanoparticles into highly ordered lattices[1,4–12]. Moreover, 3D crystals assembled entirely from DNA have been proposed and implemented with the declared goal to arrange guest molecules in predefined lattices[3,13]. This requires design schemes that provide high rigidity and sufficiently large open guest space. We here present a DNA origami-based[14,15] "tensegrity triangle" structure[16] that assembles into a 3D rhombohedral crystalline lattice. We site-specifically place 10 nm and 20 nm gold particles within the lattice, demonstrating that our crystals are spacious enough to host e.g. ribosome-sized macromolecules. We validate the accurate assembly of the DNA origami lattice itself as well as the precise incorporation of gold particles by electron microscopy and small angle X-ray scattering (SAXS) experiments. Our results show that it is possible to create DNA building blocks that assemble into lattices with customized geometry. Site-specific hosting of nano objects in the transparent DNA lattice sets the stage for metamaterial and structural biology applications.**

The spontaneous self-assembly process of rationally designed molecular building blocks enables the transition from nanoparticle to materials design[17]. One striking example is the programmed assembly of nanoparticles using DNA linkers[1,4]. Typically, spherical gold nanoparticles are modified with DNA oligonucleotides *via* covalent bonds between the gold surface and a terminal thiol residue on the DNA. The sequence-dependent hybridization of the DNA oligonucleotides mediates the formation of densely packed, 3D gold nanoparticle crystals where the connectivity and structural stability of the network is provided by the inorganic core particles. With this method, a variety of nanoparticle lattices with control over the particle-to-particle spacing and the lattice geometry have been implemented.[5–8,10,11,18,19]

An alternative approach for the site-specific positioning of nanoparticles is to use pre-assembled templates that by themselves provide structural rigidity. DNA self-assembly[13,20] allows the rational design of perfectly defined monomeric DNA nanostructures[14,15,21] and the precise tuning of monomer - monomer interaction strengths[15,22,23]. Polymerization of individual DNA motifs can result in designed macroscopic DNA crystals[3,24,25] and in large two dimensional templates that also have been employed for the 2D arrangement of guest molecules with high accuracy[26–32]. We here show the assembly of a crystalline 3D lattice purely based on DNA origami[14,15] building blocks where the lattice geometry is fully determined by the origami monomer design. Note, this approach does not rely on metal colloids and in turn, the lattice is optically transparent and only weakly scattering in the X-ray domain. We demonstrate that this type of construction allows for large unit cells and for the site-specific positioning of nano objects of up to 30 nm diameter.

For our purposes, the DNA origami building block should bear the following features: (i) high structural rigidity, (ii) polymerizability along three axes in space, and (iii) long edges to provide a large unit cell volume to host guest particles. The tensegrity triangle motif introduced by Chengde Mao & Nadrian Seeman *et al*.[3,16] fulfills the first two requirements but lacks space to accommodate large guest objects[33]. We thus built a magnified version of the tensegrity triangle with DNA origami (**Figure 1**)[34]. Each of the three edges is a 14-helix bundle with a designed length of 67 nm and a diameter of 12.5 nm. The three struts are folded from a single-stranded phage-derived scaffolding DNA (8634 nt) together with ~ 235 synthetic oligonucleotides in a temperature annealing ramp from 65 °C to 4 °C. Scaffold crossovers interconnect all struts in an over-under, over-under fashion (**Supplementary Figure S1**) predetermining the orientation of each bundle towards each other. One of the 14-helix bundles contains a “seam” where the scaffold strand does not continue through the entire edge but is closed by oligonucleotides only (inset in **Figure 1**). By choosing identical cross section, axial orientation, and staggering of the three struts, the origami monomer can polymerize *via* blunt-end stacking into a rhombohedral lattice with the unit cell parameters a = b = c = 67 nm and α = β = γ = 100 ° (≠ 90 °) (**Figure 1**). **Supplementary Figure S1** and **S2** show design details and **Supplementary Figure S3** displays different views onto the rhombohedral lattice. Note that all three upper ends (blue) match all lower ends (orange) with similar interaction strength imparting rotational freedom of the triangle around its center point (**Figure 1b**).

First, the monomeric DNA origami structures were thermally annealed starting at 65 °C (for experimental details see **Supplementary Note S1**). To avoid kinetic trapping of the struts in an undesired conformation, the oligonucleotides connecting the seam (**Supplementary Figure S2**) were injected only midway through the folding process at 52 °C. The folded triangles were purified from excess oligonucleotides by PEG precipitation before analysis by transmission electron microscopy (TEM) imaging and polymerization (**Figure 2a**). We observed that the presence of the seam oligonucleotides during the entire folding process leads to the formation of mainly deformed structures, possibly resulting from prematurely closed seams trapping the triangular struts in wrong geometries. Injection of the seam connectors at lower temperatures improved the yield of correctly folded triangles to ~ 60% (**Supplementary Figure S4**). While all misfolded objects featured three edges, they did not exhibit the designed three-fold symmetry (**Figure 2a** asterisk). Note that due to their structural similarity to the targeted design, these defective objects could not be removed during the purification steps but remained in solution during crystal growth.

To initiate the growth of the 3D lattices, “polymerization strands” were added in a ten-fold molar excess over the purified monomers. These oligonucleotides completed the formation of each of the ends of the struts and thus established shape-complementary blunt ends. The sample mix was incubated at a constant temperature of 47 °C for 90 hours and then deposited and dried on TEM grids. SEM images of origami lattices that randomly adsorbed to the grid surface show the morphology of a regular, hexagonal pattern with a center-to-center spacing between the monomers of 64 nm, which deviates slightly from the designed spacing of 67 nm and indicates flattening of the dried 3D objects on the substrate (**Figure 2b - f**). TEM images reveal the same hexagonal pattern and spacing (**Supplementary Figure S5**). Due to the obtuse interaxial angle of the building blocks, the lateral extensions along the [111] plane exceeds those of all other planes. Consequently we predominantly observe hexagonal lattices corresponding to a top view perspective of the [111] plane. For visual comparison **Supplementary Figure S6** shows 3D renderings of the designed crystals at different viewing angles. The magnified inset of the building blocks shown in **Figure 2b** confirms the over-under orientation of the struts and the left-handed chirality of the triangle as designed. Low magnification SEM images reveal the polycrystalline nature of lattice patches that are tens of micrometers in size with single domains spanning several micrometers (**Figure 2d, Supplementary Figure S5**). Close-up views display the multiple layers of the lattice and indicate its collapsed state on the dry substrate (**Figure 2e**). Of particular interest is the observation that although the defective triangles are present during the growth process they are not incorporated in the lattice patches but appear only in their periphery (**Figure 2f**). As the defective structures lack the designed symmetry, their overall binding energy does not suffice to stabilize their integration in the lattice at the elevated temperatures during lattice growth. Correctly folded monomers, instead, can replace the defective ones, which permits the self-healing growth of the origami lattices. After growth and at ambient temperatures, however, misfolded triangles can bind to any border of the lattice with just one or two connecting sites. Given the limited yield of correctly folded monomers, the observed assembly of macroscopic origami lattices indicates the effectiveness of the self-correcting processes and an overall robust lattice growth.

To demonstrate the precise placement of guest molecules in our origami lattices, we attached gold particles of different sizes at the center of each triangular origami monomer (**Figure 3a**). This choice of position preserves the symmetry upon incorporation in the lattice and therefore maintains the rhombohedral lattice type. The particle-bearing building blocks were prepared and purified as described elsewhere[35]. Consecutive lattice growth occurred under equal conditions as for the pure DNA origami lattices. Figure. 3B shows again the hexagonal pattern formed by origami triangles but this time with 10 nm or 20 nm gold nanoparticles groupings at the positions expected in this lattice orientation. Here the number of nanoparticles per group indicates the number of origami layers in the lattice. Due to the strong electron scattering properties of the gold particles only a limited penetration depth into the dense samples can be achieved and perfect hexagonal patterns can be observed for not more than a few lattice layers. A different type of nanoparticle pattern – rows of particles – results from adsorption of the lattice in an orientation different from the [111] plane (**Figure 3c, Supplementary Figure S7**). Low magnification TEM images illustrate the high quality of the particle-hosting lattices (**Figure 3d**). When placing gold nanoparticles larger than 10 nm into the monomer building blocks, we observe both in TEM and SEM images the same nanoparticle lattices as for the 10 nm particles. Clear patterns, however, only become visible at the edges of the assemblies due to the even stronger scattering of the larger particles and considerable overlap of the multiple layers (**Figure 3e, Supplementary Figure S8**).

To gain more insight into the native structure of the lattices in solution, we performed small angle X-ray scattering (SAXS) measurements for triangular origami monomers, origami lattices, and origami lattices hosting gold nanoparticles of different sizes. The scattering intensities for all samples are shown in **Figure 4a** and **Supplementary Figure S9**. Using an analytical model of three rigid cylinders each representing the 14-helix-bundles in the triangular structure we found the fitted dimensions (length=65 nm, radius=6.2 nm) to be in very good agreement with the design of the monomer. The extracted inter-helical distance of 2.8 nm matches that of previously published values for multihelical DNA bundles[36] and a constant structure factor indicates the absence of assembly into any structure of higher order. The scattering intensities of the bare origami lattice reproduce the characteristics of the triangular monomer. Additional Bragg peaks confirm the three dimensional assembly as designed. Due to the higher scattering contrast of gold, SAXS intensities of spherical gold nanoparticles placed at predefined positions within the DNA lattices predominantly show Bragg peaks indicating the lattice arrangement on top of the characteristic scattering features of spheres. **Figure 4b** shows the extracted structure factor of the samples in comparison to a model fit[37] representing a rhombohedral lattice with a lattice constant of $a = 65$ nm and $\alpha = 110$ ° for all three assemblies. More importantly, for both 10 nm and 20 nm gold nanoparticle lattices, the parameters match the ones of the pure origami lattices, which demonstrates the robustness of the lattice formation in the presence of large guest molecules. Individual peak heights and peak positions, however, differ between the pure DNA lattice and the decorated lattices due to the anisotropic nature of the triangular DNA origami monomer compared to the spherical gold particles, which is fully captured by our models (**Supplementary Note S2**). The measured unit cell of our pure DNA lattice has a volume of ~ $1.84\times10^5$ nm3, which is about 100 times larger than for previously reported DNA crystals[3] (**Supplementary Note S3**) and allows hosting guest molecules of the size of the ribosome within the origami lattice template.

In summary, we demonstrated the poly-crystalline assembly of a triangular DNA origami nanostructure into a rhombohedral lattice. With its large and rigid unit cell it can serve as a three dimensional template for the co-crystallisation of guest molecules as demonstrated here with gold nanoparticles. Further improved monomer quality and large-scale screening of crystallization conditions will potentially lead to the formation of DNA origami single-crystals that can host a wide variety of components. Importantly, the use of rigid DNA origami building blocks permits the variable positioning of guest molecules, which would allow different “guest lattices” within the same framework and even dynamically reconfigurable lattices. Combined with the optical transparency of the DNA frameworks, self-assembled metamaterials with precisely, 3D-arranged metamolecules become feasible. Furthermore our DNA lattices are almost invisible under X-ray and electron irradiation, allowing for structural analysis of any incorporated guest molecule. With a view on proteins, their 3D arrangement could open up new paths to optical super resolution-based structure analysis and CryoEM tomography.[38]

**References**.

**Acknowledgements**.

We thank Susanne Kempter for her help on TEM imaging and Philipp Altpeter on SEM imaging. We thank Alexander M. Maier, Iain MacPherson, and Florian Schüder for helpful discussions. We acknowledge support during beam time at DESY (P08, Uta Ruett) and Elettra (Austrian SAXS, Heinz Amenitsch). This work was supported by the Deutsche Forschungsgemeinschaft through the SFB grant 1032 (TPA6 and A7), the Nanosystems Initiative Munich and the European Research Council grant agreement no. 336440 for ORCA (Optical Responses Controlled by DNA Assembly). This work benefitted from SasView software, originally developed by the DANSE project under NSF award DMR-0520547. We confirm no competing financial interests.

**Author Contributions**.

T.Z., B.N. and T.L. designed the research. T.Z., P.N., and T.L designed the DNA structure. T.Z., C.H., A.H.-J. prepared the assemblies. C.H., S.F., K.F., and B.N. performed the SAXS measurements. S.F., K.F., C.H. and B.N. analyzed the SAXS data. T.Z., S.F, K.F., and C.H. prepared the figures. C.H., T.Z., B.N. and T.L wrote the manuscript. All authors discussed the results and edited the manuscript.

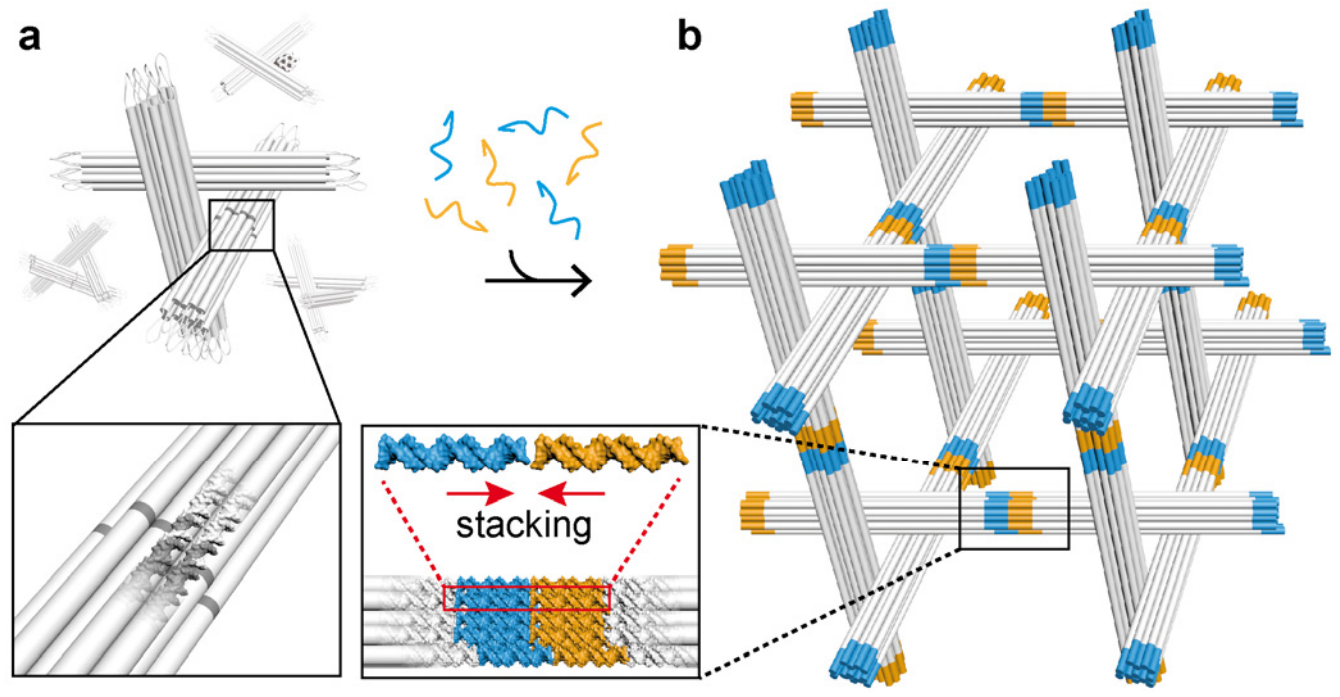

**Figure 1.** Schematic illustration of the triangular DNA origami building block design and assembly. **a**, The three 14-helix bundles of equal lengths are interconnected at defined positions and form a constrained triangular structure. A scaffold seam region is bridged by “connection” oligonucleotides (inset). **b**, Addition of oligonucleotides completing the struts results in the formation of self-matching shape complementary blunt ends. The three-fold symmetry facilitates the polymerization of the triangular monomers into a rhombohedral lattice *via* stacking interactions (inset). Details on the assembly process and additional 3D views are shown in **Supplementary Figure S1** & **S2**.

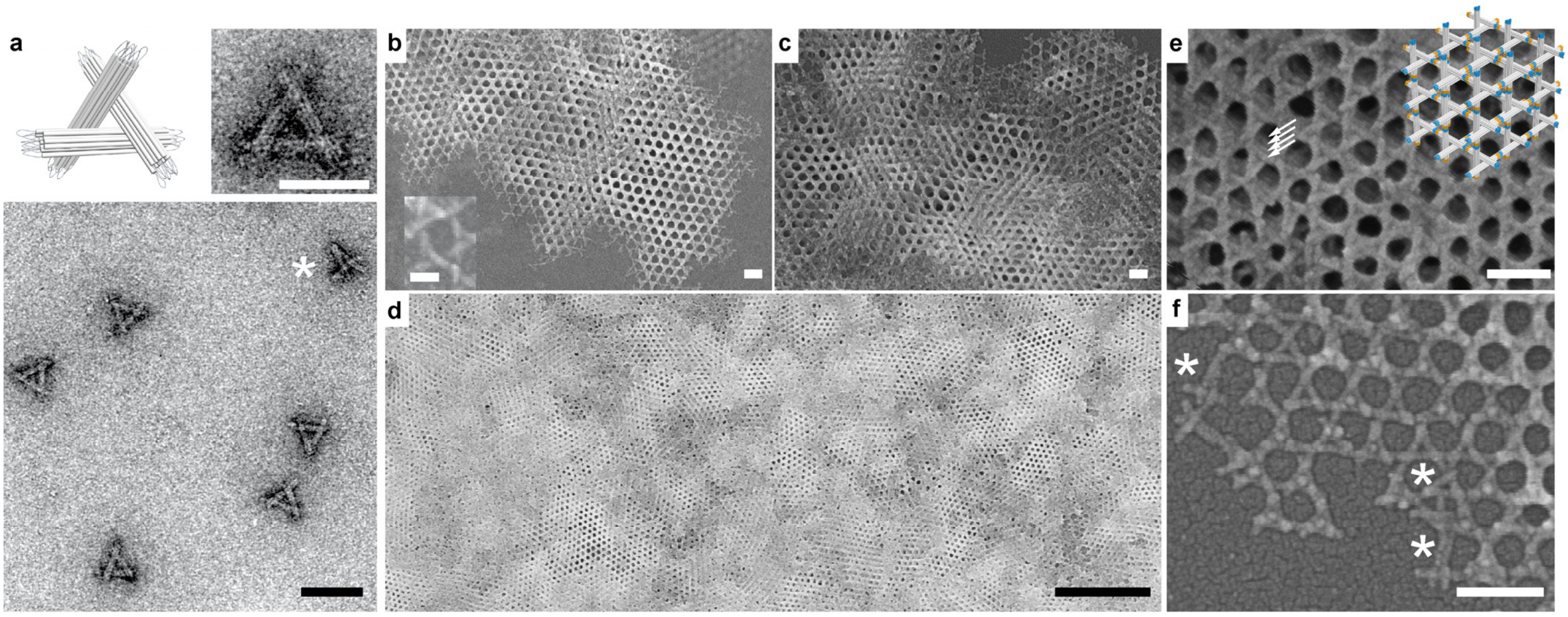

**Figure 2.** DNA origami lattices. **a,** TEM images of purified triangular DNA origami structures after purification and before addition of polymerization oligonucleotides that would initiate lattice growth. The asterisk points out a misfolded monomer. **b-f,** SEM images of DNA origami lattices. The inset in panel **b** exemplifies the left-handed over-under design. Panel **c** reveals the three-dimensionality of the assemblies and their polycrystallinity becomes apparent in the wide-field view shown in panel **d**. Although the lattices collapse on the imaging substrates during drying, the multiple layers and the original geometry can be inferred in a magnified view (white arrows, panel **e**). **f,** SEM image showing the border of a lattice with asterisks indicating defective structures that were expelled from the lattice during the growth process and can only bind later at lower temperatures to the periphery. Scale bars in **a, b, c, e**, and **f**: 100 nm, **d**: 1 µm, insets in **a** and **b**: 50 nm.

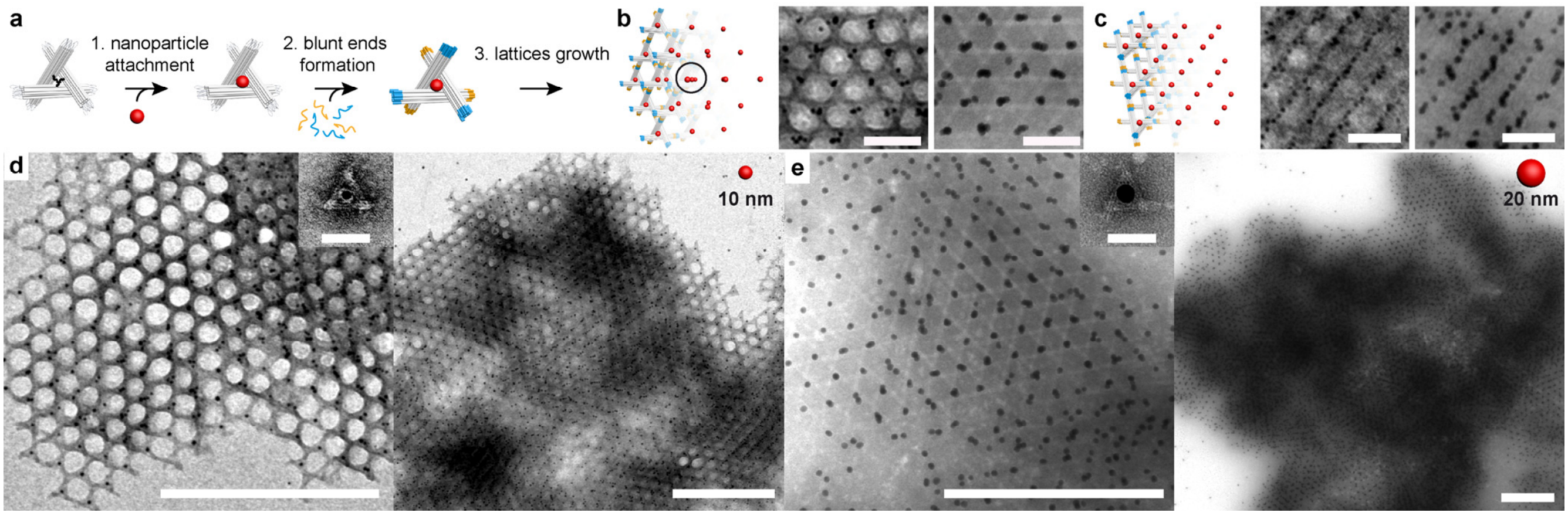

**Figure 3.** Hosting of gold nanoparticles in DNA origami lattices. **a,** Workflow to prepare gold nanopaticle lattices: 1) Folded and purified DNA origami triangles are incubated with gold particles overnight and then purified from excess particles. 2) Addition of the polymerization oligonucleotides to initiate the lattice growth process. 3) Incubation at 47°C for 3-4 days yields assembled host-guest lattices. **b, c,** Model views and TEM images of DNA origami lattice hosting 10 (left) or 20 nm (right) gold nanoparticles. The number of particles per grouping indicates the number of lattice layers overlapping at the respective points (black circle in **b**). **d, e,** Wide-field TEM images of origami lattices hosting 10 nm (**d**) and 20 nm (**e**) gold nanoparticles. Scale bars in **b** and **c**: 100 nm; **d** and **e** 500 nm, insets: 50 nm.

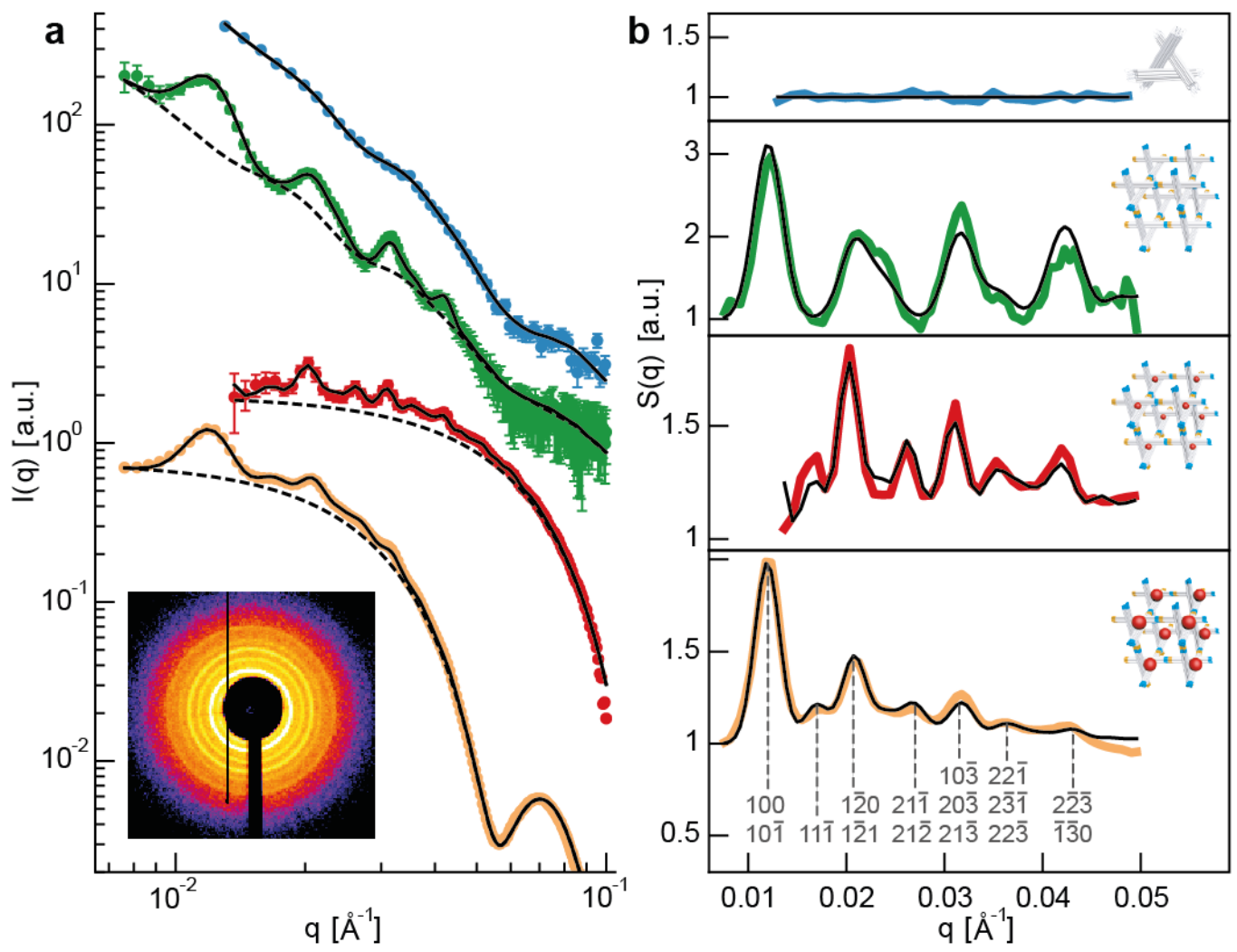

**Figure 4. a,** SAXS intensities of triangular DNA origami monomers (blue), origami lattices (green), and origami lattices hosting 10 nm (red) and 20 nm (orange) gold nanoparticles (vertically offset). Solid black lines: Model fits of total intensity. Dotted black lines: Simulated form factor intensities neglecting the lattice structure factors. Inset: SAXS pattern of 10 nm gold-decorated sample. **b,** The monomer SAXS intensity (blue) divided by the fit yields a constant, indicating the absence of a lattice. Lattice structure factors from SAXS data (green, red, yellow) obtained by subtraction of background and division by the form factor intensity (vertically offset). Black lines: Model fits assuming a rhombohedral unit cell (a = 65 nm, *alpha* = 110°). Dashed lines with Miller indices label selected Bragg peaks.

**Supplementary Notes S1: Materials and Methods**

<u>Design and formation of DNA origami.</u>

*Design*. The triangular tensegrity origami was designed using caDNAno (design schematics in **Supplementary Figure S1, S2**)[1]. The structure consists of three 14 helix bundles (14HBs) packed on a honeycomb lattice with a diameter of about 12.5 nm. These three 14HBs were designed to be of the same length (199 bp) and display self-matching shape complementary blunt ends. The 14HBs were interconnected at selected positions by forced crossovers with three bases of scaffold spacers. In order to avoid topological traps of the struts in undesired geometries, a "seam" was introduced into of the 14HBs. Here the scaffold does not run from one end of the strut to the other but loops back in the middle of the strut for each pair of helices (**Supplementary Figure S1b**). The seam is closed by staple oligonucleotides. Groups of staple strands were divided into "connection" oligonucleotides (closing the seam), "polymerization" oligonucleotides (completing the ends of all struts and thus enabling blunt end stacking), handle oligonucleotides (for capturing the gold nanoparticles) and "core" oligonucleotides (all other strands).

*Folding and Purification*. DNA origami structures were prepared by mixing core and handles staples (100 nM each, MWG Eurofins), and the circular DNA scaffold strand p8634 (12.5 nM, produced in house) in 1x TE-$Mg^{2+}$ buffer (10 mM Tris, 1 mM EDTA, 18 mM $MgCl_2$). The mixture was thermally annealed from 65 °C to 4 °C over 35 h (15 min at 65 °C, cooling to 58 °C with a cooling rate of −1 °C per 5 min, 58 °C to 35 °C with rate of −1 °C per 1 h, and from 35 °C to 4 °C with rate of −1 °C per 5 min). Connection staples were injected into the folding mixture during the annealing process at 52 °C (**Supplementary Figure S4**). Subsequently the folded DNA nanostructures were purified from excess DNA staples by agarose gel electrophoresis stained with 1x Sybr Safe (1 % agarose in1x TAE 11 mM $MgCl_2$ buffer; 6.5 V/cm for 2 h) or by polyethylene glycol (PEG) precipitation([2], [3]). For agarose gel purification, samples were run in 1 % agarose gels containing 1x Tris-acetate buffer (10 mM Tris, 10 mM acetic acid), 11 mM $MgCl_2$ and 1x SybrSafe (Thermo Fisher Scientific). All gels were cooled in ice water baths. Samples were separated at 6.5 V/cm for 2 h following excision of the bands and recovery of the products by squeezing the band between two glass slides and collecting the resulting liquid droplet with a pipet. For PEG precipitation, equal volumes of 2x PEG buffer (15 % (w/v) PEG-8000, 2x TE, 500 mM NaCl, 20 mM $Mg^{2+}$) and unpurified folding solution were mixed and centrifuged for 30 min at 16,000 rcf. The resulting pellet was re-suspended in 1x TAE, 11 mM $MgCl_2$ buffer and subsequently shaken at 650 rpm, 30 °C for 24 h in order to re-disperse the origami structures.

<u>Formation of DNA origami-gold nanoparticle conjugates.</u>

*Functionalization of gold nanoparticles with DNA.* Gold nanoparticles (AuNPs) of 10 nm, 20 nm and 30 nm (BBI International) were functionalized with 5'-thiolated 19T single-stranded DNA (Biomers) following published methods[4]. Briefly, TCEP (tris(2-carboxyethyl)phosphine) treated thiolated DNA was added in excess (200x molar excess for 10 nm AuNPs, 800x for 20 nm AuNPs, and 1800x for 30 nm AuNPs) to AuNPs. The mixture was incubated at room temperature for 24 hours before slowly increasing the salt concentration to 500 mM by addition of 1 M NaCl over a period of 6 h. DNA-modified gold nanoparticles were then purified using Amicon centrifugation filters (100K MW cut-off).

*Conjugation of gold nanoparticles to DNA origami.* Six staple oligonucleotides at the center of the triangular origami structure were extended from the structure with 19A bases serving as handles for attachment of the gold nanoparticles (one nanoparticle per origami nanostructure). An 8x molar excess of DNA-AuNPs was added to the DNA origami structures and incubated overnight. The resulting DNA origami-gold nanoparticle conjugates were purified from excessive gold nanoparticles in 1 % agarose gels containing 1x Tris-acetate, 11 mM $MgCl_2$ buffer, cooled in an ice-bath. Samples were separated at 7 V/cm for 1.5 h. Bands were excised from the gel and recovered as described previously.

Polymerization of DNA origami into 3D lattice.

*Polymerization.* Polymerization oligonucleotides were mixed at 10x molar excess with purified bare triangular origami structures or with purified structures carrying gold nanoparticles. The buffer was then brought to 1x TAE and 15 mM $MgCl_2$. The polymerization mixture was incubated at a constant temperature of 47 °C for 3-4 days in a thermocycler. Owing to the different purification procedures and varying purification yields, the starting concentration of origami monomer for pure DNA origami lattice growth was ~ 25 nM (PEG purified), while for lattices with gold particles the starting concentration of the gold-carrying nanoparticles ranged between 1 nM and 3 nM (gel purified).

Characterization techniques.

*TEM.* TEM imaging of DNA origami lattices was carried out using a JEM-1011 transmission electron microscope (JEOL) operating at 80 or 100 kV. For sample preparation 10 µL of polymerized DNA origami structures were deposited on glow-discharged TEM grids (formvar/carbon-coated, 300 mesh Cu; TED Pella, Inc; prod no. 01753 - f) for 1 h. For pure origami lattices and origami lattices containing 10 or 20 nm gold nanoparticles, grids were furthermore quickly washed once with 0.1 ‰ uranyl acetate solution (5 µL) and immediately afterwards stained with 0.1 ‰ uranyl acetate solution (5 µL) for 10 s. For origami lattices hosting 30 nm gold nanoparticles and the lattices for taking SEM images, grids were washed two times with water for 2 s.

*SEM.* The TEM grids were directly used for SEM imaging after 10 s sputtering using an Edwards Sputtercoater S150B 1990. The sputter target contained 60 % gold and 40 % palladium. Process parameters used for sputtering were 7 mbar Ar, 1.1 kV, 35 mA. 10 s sputtering results in the deposition of layer of gold/palladium with a thickness of a few nm. The Au/Pd deposited TEM grids were directly fixed on the sample holder with carbon tape for SEM imaging with a Carl Zeiss LEO DSM 982 GEMINI (containing a source of thermal field emitting (TFE) cathode (1997) and a detector of LEO High Efficiency In-Lense Secondary Electrons). Beam parameters for taking imaging were set as 5 kV acceleration voltage and 30 µm aperture.

*SAXS.* The SAXS data were measured at three different sources. All sample to detector distances and beam centers were calibrated with silver behenate. The scattering data of the monomer was measured at an in-house X-ray source, which is described in detail in the literature[5]. The data of the DNA origami lattice decorated with 10 nm gold nanoparticles and the undecorated DNA origami lattices shown in **Fig. S9** were measured at the beamline P08 at PETRA III (DESY) in Hamburg. A Perkin Elmer flat panel XRD 1621 with 2048 x 2048 pixels with 200 µm size served as detector. The solutions of polymerized sample were loaded in 2 mm quartz capillaries. The measurement was carried out at 20 keV in order to avoid radiation damage of the sample. The data of the DNA origami lattice decorated with 20 nm gold nanoparticles and the undecorated DNA origami lattice shown in Figure 4 were measured at the SAXS beamline at ELETTRA in Trieste The solutions of polymerized sample were loaded in 1 mm quartz capillaries and measured at 8 keV X-ray energy. A Dectris Pilatus 3 1M CMOS detector with 981 x 1043 pixels with 172 µm pixel size served as detector.

**Supplementary Notes S2: SAXS data analysis.**

The scattering intensity is shown as a function of scattering vector $q = \frac{4\pi}{\lambda}\sin\theta$ where $\lambda$ is the wavelength of the monochromatic X-ray radiation and $\theta$ is one half of the scattering angle.

The scattering intensity of the DNA origami monomer was modeled as $I(q) = P(q) + bg(q)$. To obtain the isotropic form factor intensity $P(q)$ the analytical expression of the form factor of a rigid cylinder representing a 14 helix bundle was added up once for each bundle. The different axis directions of the bundles inside a monomer were taken into account by a coordinate transform and the spatial offset of their centers by phase factors. Orientational averaging was carried out by numeric integration.

Following the procedure described in *Ref.* [6], the scattering intensities of the lattice assemblies were modeled as

$$I(q) = P(q) \cdot S(q) + bg(q)$$

where $P(q)$ is the isotropic form factor intensity of the particle in the unit cell. For the origami lattice, $P(q)$ equals the monomer form factor intensity described above. For the decorated lattices, the gold particle form factor intensity was found to dominate the total intensity due to the higher scattering contrast of gold compared to DNA. Therefore, the isotropic form factor intensity of a sphere was used as $P(q)$ in these cases. $S(q)$ is the structure factor and $bg(q)$ a background contribution consisting of a constant term and a $q$-dependent term accounting for excess oligonucleotides in the sample solution. The latter term was modeled by a Debye model representing Gaussian polymer chains.

The structure factor $S(q)$ was computed as

$$S(q) = \frac{cZ_0(q)}{P(q)} G(q) + 1 - \beta(q) G(q).$$

$c$ $c$ is a scaling constant. $Z_0(q)$ is the lattice factor, representing the structural scattering in terms of Bragg peaks.

$$Z_0(q) = \frac{1}{q^2} \sum_{hkl}^{m_{hkl}} \left| \sum_{\vec{r}_j} F_j(\vec{q}_{hkl}) \cdot e^{-i\vec{q}_{hkl} \cdot \vec{r}_j} \right|^2 \cdot L(q - q_{hkl})$$

$m_{hkl} = 7$ is the maximum absolute value of the Miller indices $hkl$ which were considered. $F_j(\vec{q}_{hkl})$ is the form factor of the particle at position $\vec{r}_j$ in the unit cell, taking into account its orientation. $L(q - q_{hkl})$ is a Gaussian peak-shape function. $\vec{q}_{hkl}$ is the scattering vector corresponding to a Bragg peak with Miller indices $hkl$. In the case of the origami lattice it was computed as $\vec{q}_{hkl} = h\vec{b}_1 + k\vec{b}_2 + l\vec{b}_3$ by calculating the reciprocal lattice vectors $\vec{b}_1, \vec{b}_2, \vec{b}_3$ to the rhombohedral real lattice vectors spanned by the 14 helix bundles. In the case of the gold-decorated lattices the formula

$$q_{hkl} = \frac{2\pi}{a} \sqrt{\frac{(h^2 + k^2 + l^2)\sin^2\alpha + 2(hk + kl + lh)(\cos^2\alpha - \cos\alpha)}{1 - 3\cos^2\alpha + 2\cos^3\alpha}}$$

for a rhombohedral real space lattice with lattice constant $a$ and unit cell angle $\alpha$ was used.

For anisotropic particles the form factor intensity which enters the lattice factor via $|F(\vec{q}_{hkl})|^2$ differs from the isotropic form factor intensity $P(q)$. The orientation of the particle with respect to the unit cell therefore becomes important. This needs to be considered for the origami lattice, where the form factor intensity of the anisotropic triangular monomer enters into the total scattering. This orientation effect on the peak heights of the structure factor is illustrated in **Supplementary Figure S9 c, d** and **e**. The fitted peak heights and positions agree with the measured scattering intensity when taking into account the orientation of the DNA origami monomer in the unit cell. If the orientation is neglected and $|F(\vec{q}_{hkl})|^2$ is replaced by the isotropic form factor intensity $P(q)$ the peak heights are not correctly reproduced and additional peaks appear in the simulated curve.

$G(q) = e^{-\sigma_d^2 a^2 q^2}$ models lattice disorder which leads to a damping of higher order peaks and a diffuse baseline in the structure factor. $\sigma_d$ is the r.m.s. displacement of particle position normalized to the lattice constant $a$. $\beta(q) = e^{-(\sigma_R \cdot R)^2 q^2}$ accounts for particle size polydispersity. $(\sigma_R \cdot R)$ is the r.m.s. variation of the size of the particles, i.e. the origami monomers in the case of the origami lattice and the gold particles in the case of the decorated lattices.

All models were written as C code in the software package SasView (http://www.sasview.org/). Fit parameters for the total scattering intensity $I(q)$ were obtained by running the software-internal population-based DREAM algorithm. Already during fitting all model intensity curves were smeared with a pinhole function adapted to the respective instrumental resolution of the data.

All experimental structure factors shown were obtained by subtracting the background $bg(q)$ from the scattering intensity data and dividing the result by the form factor intensity $P(q)$. Both $bg(q)$ and $P(q)$ as well as the modeled structure factor $S(q)$ shown for comparison were obtained from one fit to the total scattering intensity $I(q)$.

For both undecorated and decorated origami lattices (**Supplementary Figure** S9 **f-i**) the sharpest and highest Bragg peaks are observed in the samples assembled at 47 °C. Sharp and high peaks indicate the presence of large crystallites and corroborate the observations from TEM images. Therefore this incubation temperature was chosen for all following studies and used for all crystalline samples presented in this work. The Scherrer equation

$$L_c = \frac{2\pi K}{\Delta q}$$

gives an estimate of the crystallite size $L_c$, where K is a shape factor of the order of 1 and $\Delta q$ is the full width at half maximum of a diffraction peak. The dominating contribution to the peak width in all SAXS data shown is from the instrument resolution function $\Delta q \approx 1 \cdot 10^{-3}$ Å$^{-1}$ which was calculated from the beam size and divergence as well as the detector pixel size. Applying Scherrer equation yields a lower limit to the linear crystallite size of at least 6300 Å $\approx 10$ unit cells.

**Supplementary Notes S3: Cavity size calculation.**

The cavity size of the unit cell should be the unit cell volume minus the bundle volume. We used the formula (E1) for volume calculation. For each 14-helix bundle, it distributes its volume to four neighboring unit cells of a volume given by formula (E2). The cavity size calculated for the triangular DNA origami crystal (bundle radius *r* = 6.2 nm, rhombohedral unit cell constants a= 65 nm, *α = 110*°) is 183616 nm$^3$ which is about 115 times larger than the one previously reported ([7]) for a DNA duplex crystal (duplex radius *r* =1.25 nm; rhombohedral unit cell constants *a* = 13.49 nm, *α =110.9*°) of 1584 nm$^3$ .

$$v = a^3\sqrt{1 - 3\cos^2\alpha + 2\cos^3\alpha} - bundle\ volume \qquad \text{(E1)}$$

$$bundle\ volume = (12 \times \frac{1}{4})(\pi \times r^2)a \qquad \text{(E2)}$$

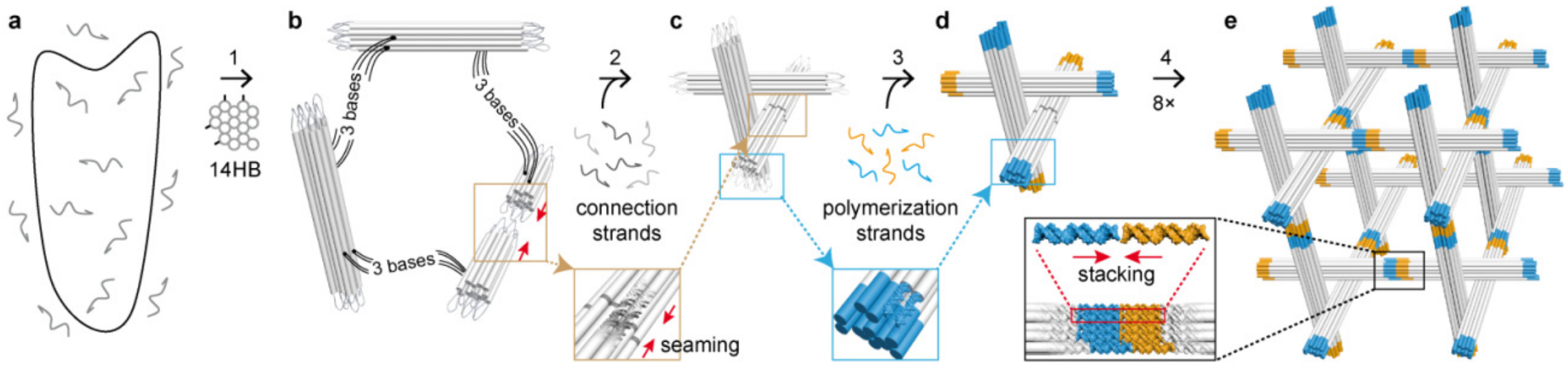

**Supplementary Figure S1.** Structural details of the triangular DNA origami design and the sequential workflow towards DNA origami lattices. **a**, The circular scaffold strand and the core staple strands (without connection oligonucleotides and polymerization oligonucleotides, caDNAno layout in **Supplementary Figure S2**) are annealed in step 1 to form the three 14-helix bundles (14HBs) of the triangular DNA tensegrity structure. The resulting honeycomb design of the 14HBs and the positions where the scaffold crosses over between the struts (black lines) are shown in the cross section image. **b**, Schematic of the 3 nucleotide-long scaffold spacers connecting the three 14HBs. The selected connection positions and the shortness of the scaffold loops results in a structurally self-restricting origami triangle structure. The inset depicts the scaffold seam in one of the 14HBs. This seam is closed in step 2 by the addition of connection strands during the folding cycle. This two-step process favors correct geometries (over-under, over-under arrangement) over misconnected triangles. **c**, Correctly folded DNA origami monomer. The addition of polymerization strands in step 3 results in the completion of the struts' ends and the formation of dsDNA blunt ends (inset). **d**, As the DNA origami monomer displays a three-fold symmetry, any blue end can interact with any orange end (inset). Thus the blunt end stacking of the shape complementary ends leads to polymerization of the monomers in three out-of-plane directions and to the growth of a rhombohedral lattice (step 4). **e**, The resulting unit cell shares eight triangular DNA origami monomers**.**

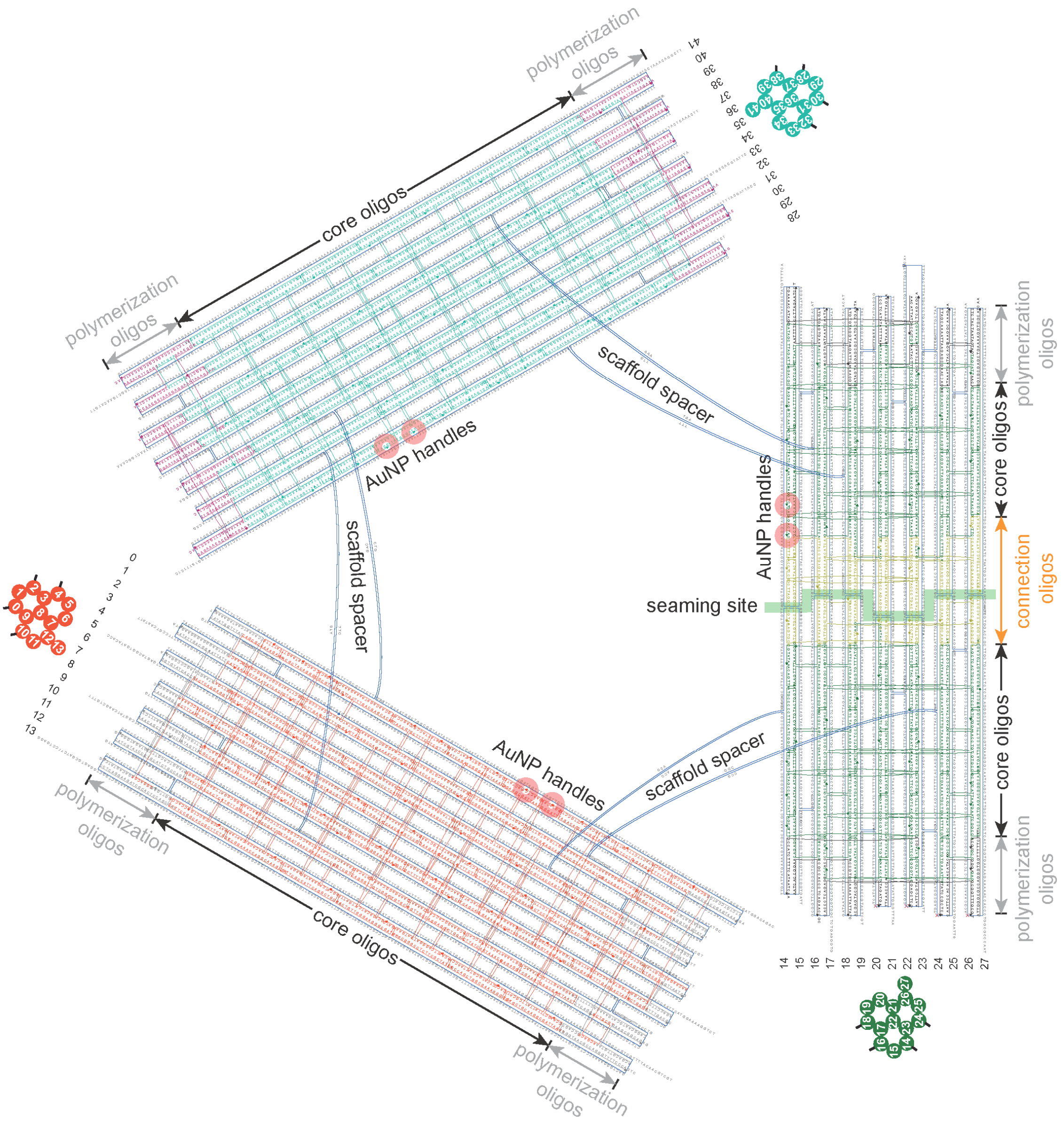

**Supplementary Figure S2.** Rearranged caDNAno layout of the triangular DNA origami monomer. The staple oligonucleotides are sorted into different groups: core oligonucleotides, connection oligonucleotides, polymerization oligonucleotides, and handle oligonucleotides. Six handles (3'-end) for capturing together one gold nanoparticle (AuNP) are highlighted with red circles. The seam is highlighted with a bold green zigzag line.

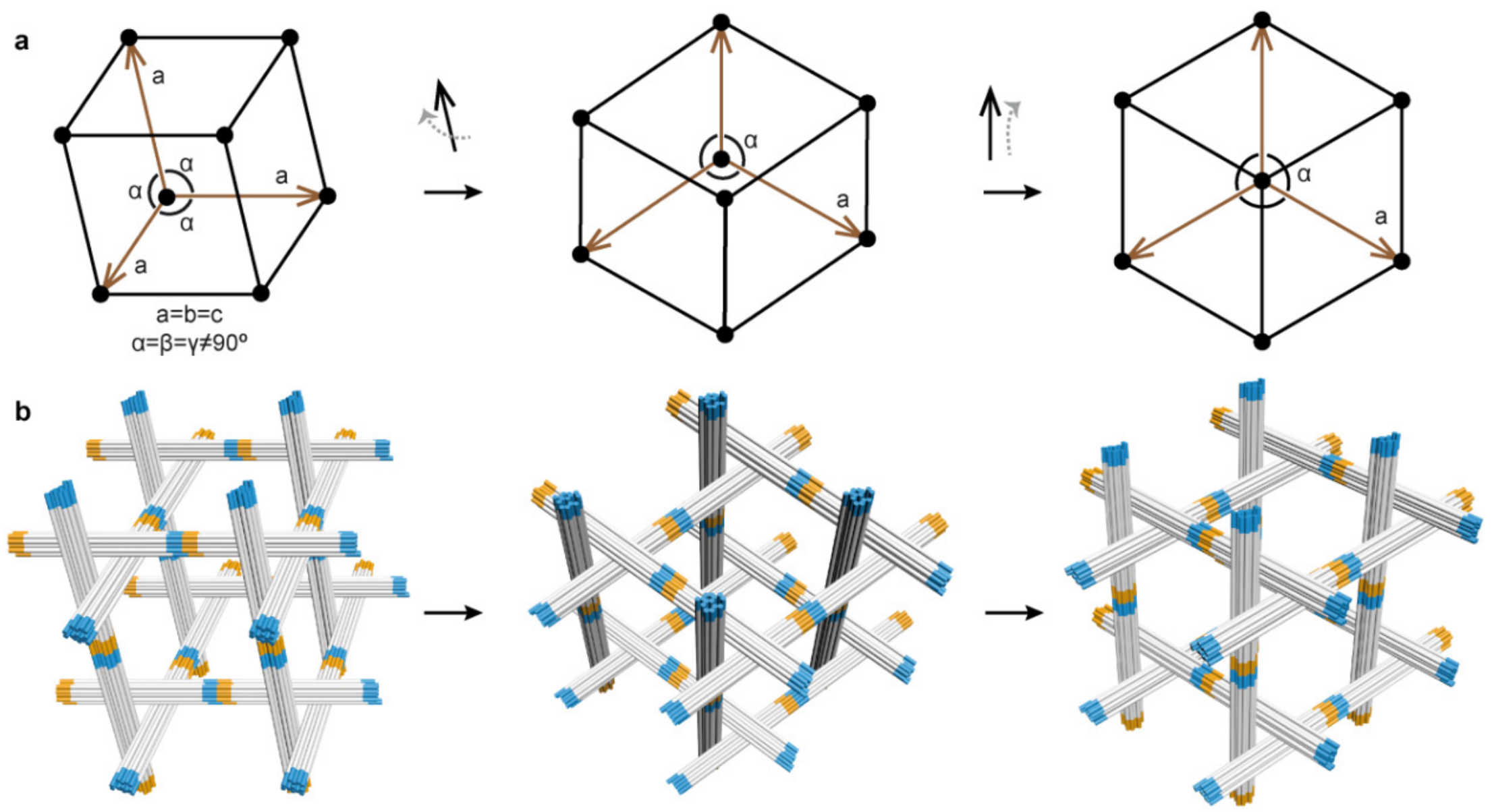

**Supplementary Figure S3. a**, Illustration of the rhombohedral unit cell with lattice constants a = b = c = 67 nm and α = β = γ = 100° (≠ 90°), matching the designed parameters. A step-by-step rotation exemplifies the appearance of a hexagonal pattern ({111} plane) with three-fold symmetry in the unit cell of the rhombohedral lattice. **b**, Schematic of the corresponding rhombohedral lattice unit cell consisting of eight triangular DNA origami structures.

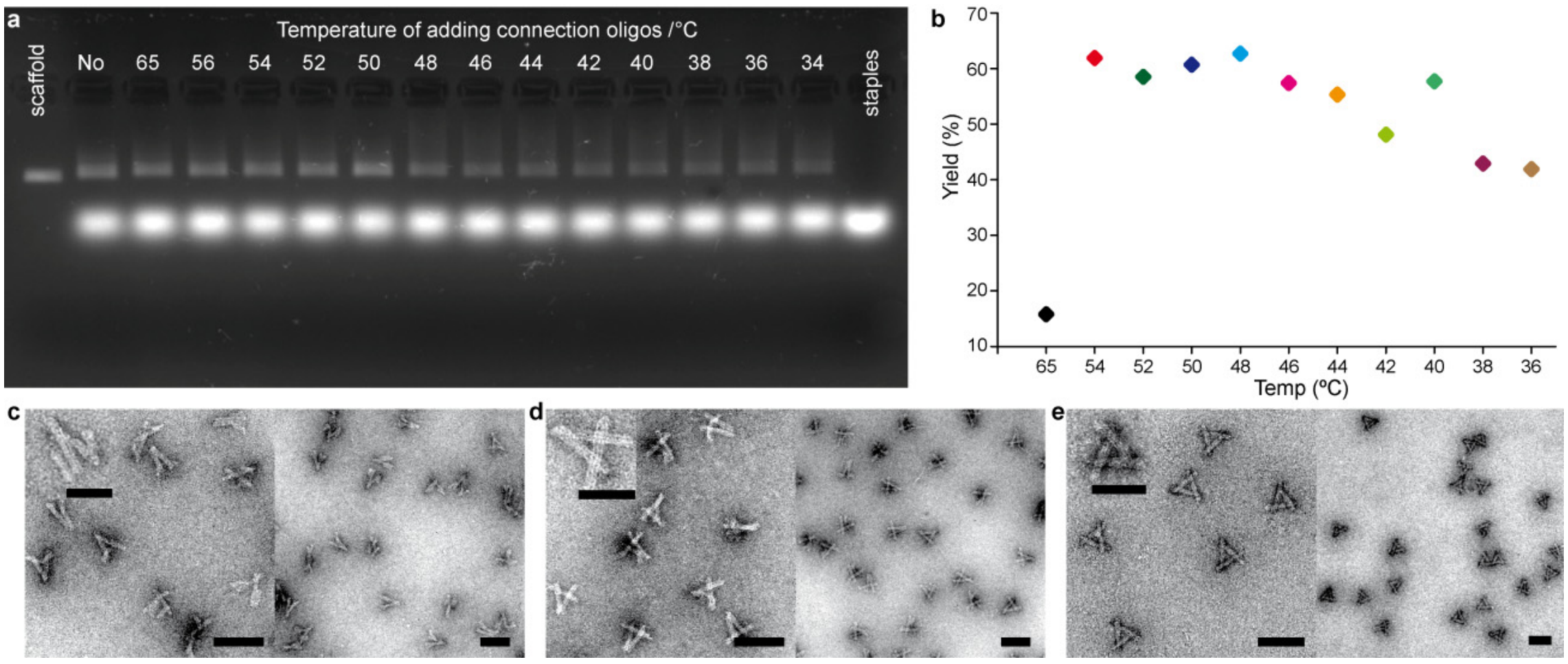

**Supplementary Figure S4. a**, Agarose gel electrophoresis (0.7 %, 11 mM $Mg^{2+}$, 1× TAE pH 8.0) of DNA origami triangles with connection staples added at different temperatures. No: no connection strands added; 65: DNA origami folded with connection oligonucleotides added from the very beginning (one-pot reaction); 56-34: the temperature in °C at which connection strands were added. Correctly folded structures cannot be separated from defective ones *via* agarose gel electrophoresis. **b**, Analysis by TEM imaging revealed that addition of the connection oligonucleotides in a temperature range from 54 to 48 °C resulted in ~ 60% of correctly folded structures. This yield turned out to be sufficient to grow lattices and it is considerably higher than for foldings, where the connection staples were added from the beginning resulting in 15 % correctly folded structures. **c**-**e**, TEM images of triangular DNA origami structures folded c, without connection oligonucleotides, d, in a one-pot reaction,and e, with connection oligonucleotides added at 52°C. Scale bar of inset: 50 nm, all others: 100 nm.

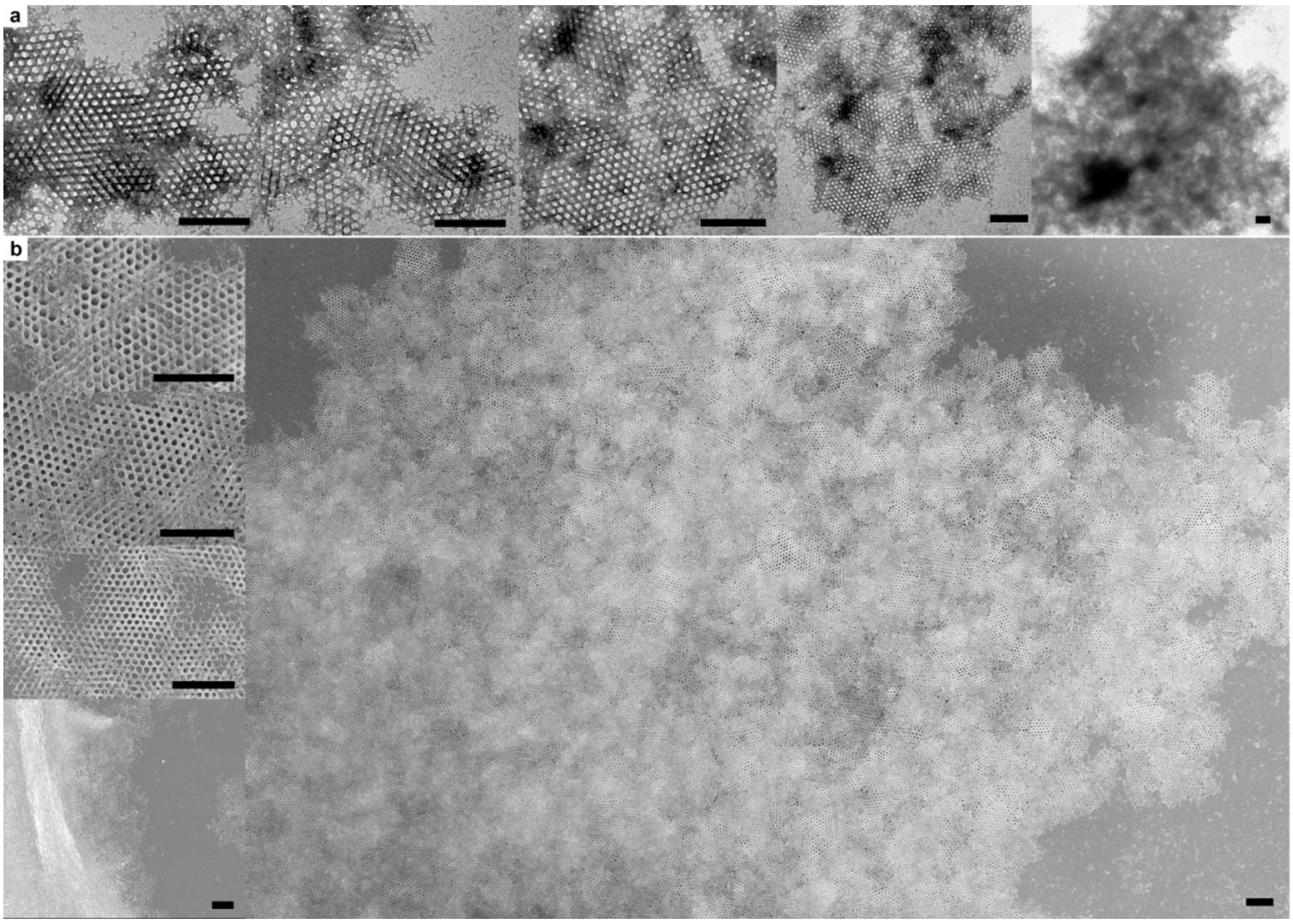

**Supplementary Figure S5. a**, TEM images of the DNA origami lattices. The hexagonal pattern characteristic for the view on the [111] plane can be observed particularly well in lattices consisting of only a few layers. Thicker lattices are barely penetrated by the electron beam. **b**, SEM images of the DNA origami lattices displaying the same hexagonal pattern. The lowest small image on the left shows a large lattice patch that buckled during the adsorption on the substrate. The large low magnification SEM image reveals the polycrystalline nature of the DNA origami lattices. Scale bars: 500 nm.

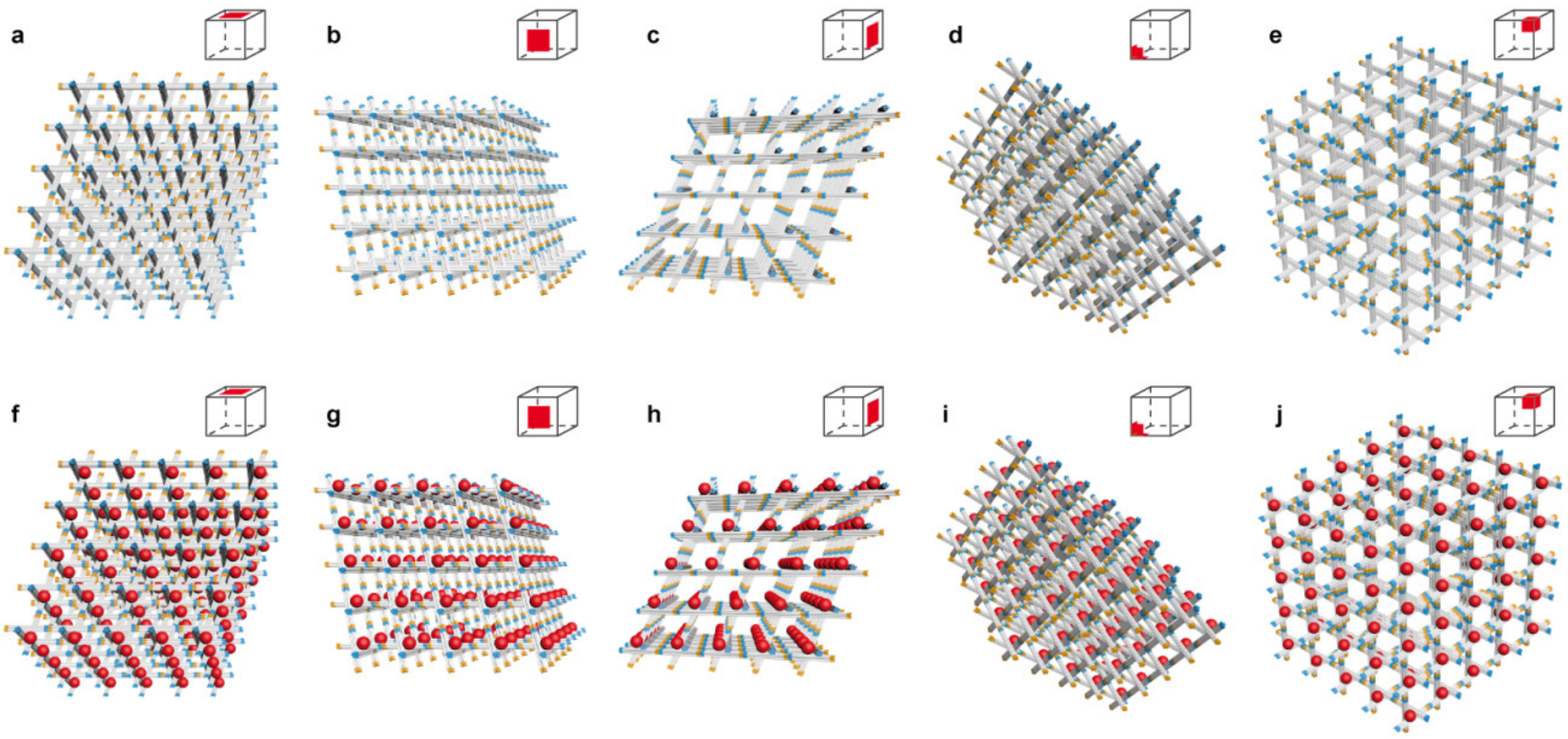

**Supplementary Figure S6.** 3D rendering of the DNA origami lattice (**a**-**e**) and the lattice hosting 20 nm gold nanoparticles (**f**-**j**) viewed from different perspectives.

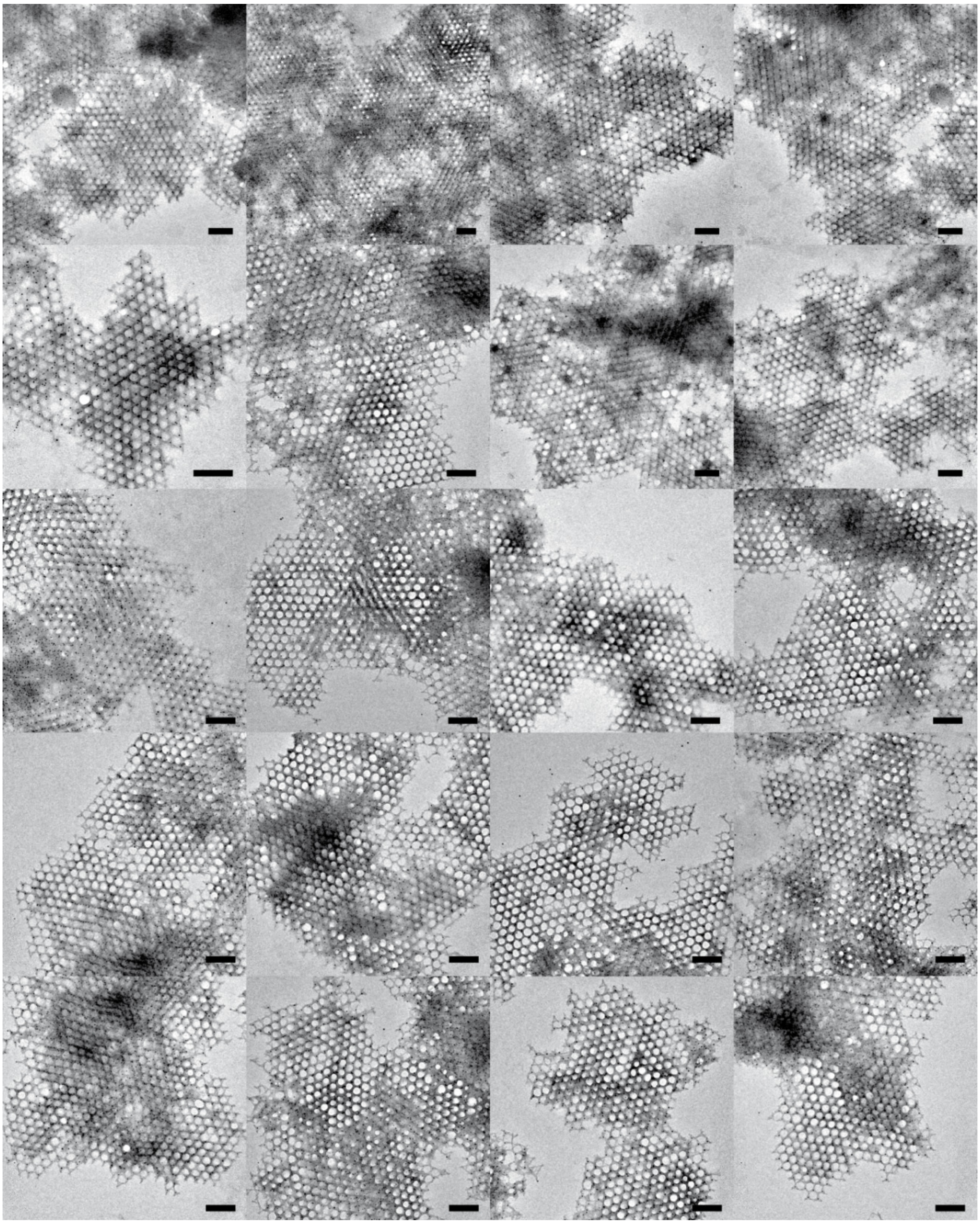

**Supplementary Figure S7.** TEM images of the DNA origami lattices hosting 10 nm gold nanoparticles. Scale bars: 200 nm.

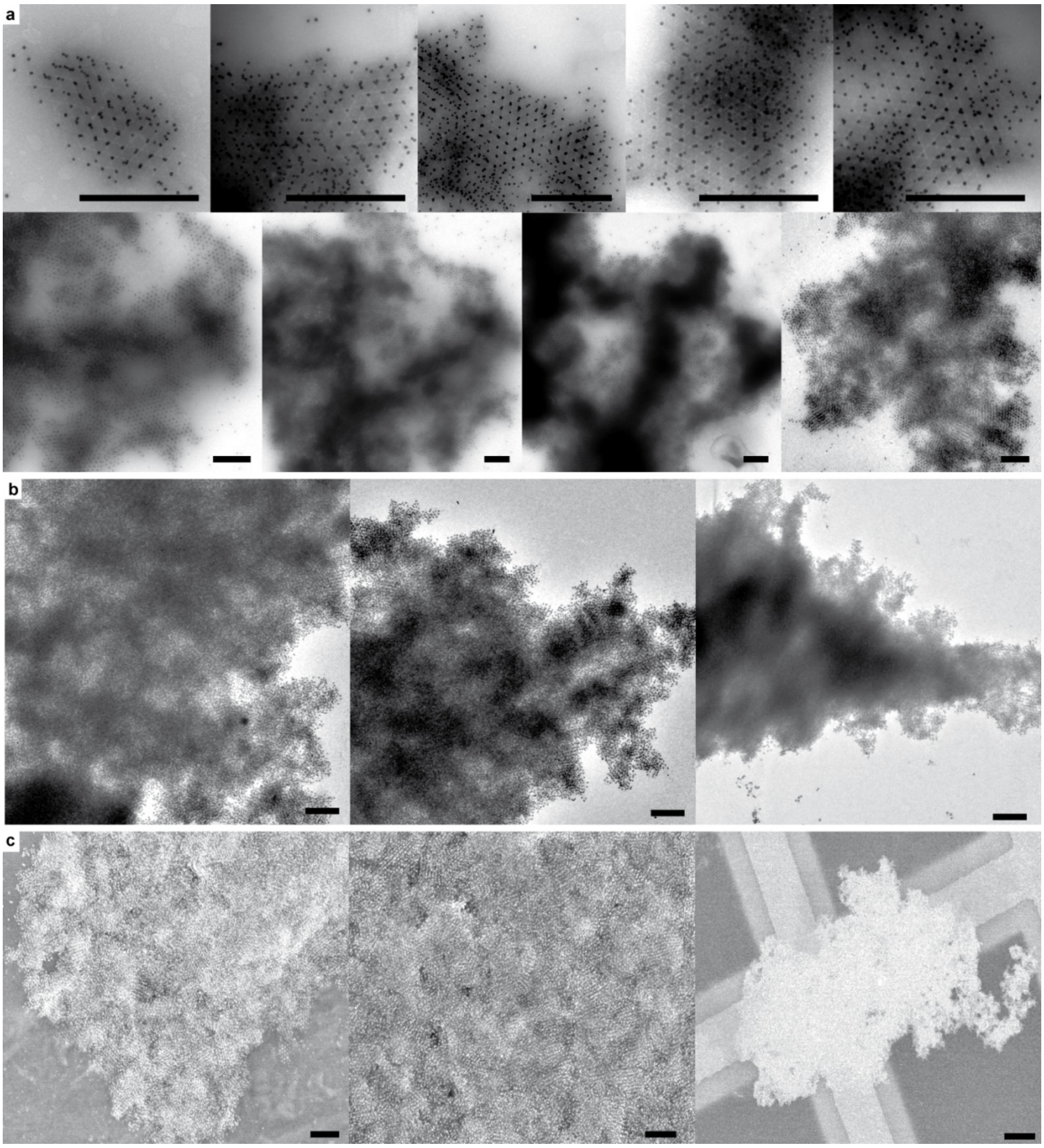

**Supplementary Figure S8.** TEM images of DNA origami lattices hosting 20 nm (**a**) and 30 nm (**b**) gold nanoparticles. **c**, SEM images of DNA origami lattices hosting 20 nm gold nanoparticles exhibit the same poly-crystallinity as the templates. Except for the bottom right low magnification image where the scale bar is 10 µm all other scale bars: 500 nm.

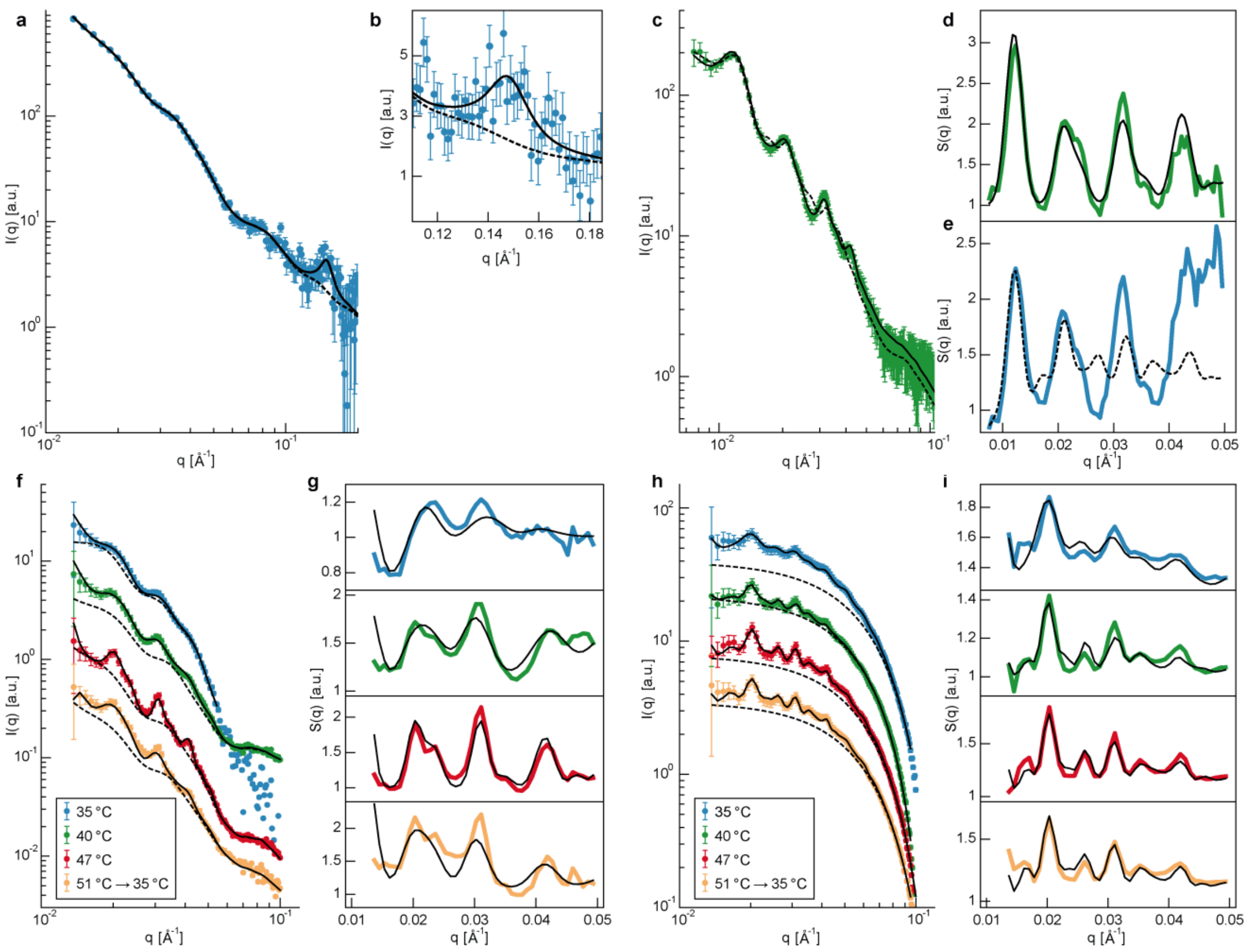

**Supplementary Figure S9. a**, SAXS intensity of triangular DNA origami monomers dispersed in solution recorded with an in-house source. Model fit to the total intensity including a Lorentzian peak accounting for the regular honeycomb arrangement of DNA double-helical domains inside the 14-helix bundles (solid line) and model fit ignoring this feature (dotted line). **b**, Zoom into the Lorentzian peak. Using the procedure described in *Ref.* [8] we obtain an interhelical distance of 2.8 nm. This is in reasonable agreement with the value of 2.5 nm that was obtained for a honeycomb design in the previous study with an accelerator source. **c**, SAXS intensity of the DNA origami lattices dispersed in solution. Solid line: Complete model fit taking into account the orientation of the triangular monomer with respect to the unit cell. Dashed line: Incomplete model fit neglecting this orientation. **d**, Extracted structure factor from the complete model fit (green line) and model of the complete structure factor assuming a rhombohedral unit cell (black line). Peak positions and heights match. **e**, Extracted structure factor from the incomplete model fit (blue line) and model of the incomplete structure factor assuming a rhombohedral unit cell (black dotted line). Peak heights do not match the data and additional peaks appear where minima are observed in the data. **f-i**, SAXS intensity of origami lattices (**f**) and origami lattices decorated with 10 nm gold particles (h) dispersed in solution that were polymerized at different temperatures (blue, green, red) and with an annealing ramp from 51 °C to 35 °C (orange), vertically offset. Solid black lines: Model fits of the total intensity. Dashed black lines: Simulated form factor intensity, neglecting the lattice structure factor. Lattice structure factors from SAXS data (**g** & **i** from **f** & **h** respectively) were obtained by subtraction of the background and division by the form factor intensity. Black lines: Model fits assuming a rhombohedral unit cell.